\documentclass[aps,pre,twocolumn,groupedaddress,amsmath,amssymb]{revtex4-1}
\usepackage{graphicx}  
\usepackage{dcolumn}   
\usepackage{bm}        
\usepackage{verbatim}   

\begin{document}

\title{Perturbative Thermodynamic Geometry of Nonextensive Ideal Classical, Bose and Fermi Gases }
\author {Hosein Mohammadzadeh}
\email{mohammadzadeh@uma.ac.ir}
\affiliation{Department of Physics, University of Mohaghegh Ardabili, P.O. Box 179, Ardabil, Iran}
\author {Fereshteh Adli}
\affiliation{Department of Physics, University of Mohaghegh Ardabili, P.O. Box 179, Ardabil, Iran}
\author {Sahereh Nouri}
\affiliation{Department of Physics, University of Mohaghegh Ardabili, P.O. Box 179, Ardabil, Iran}


\begin{abstract}
We investigate perturbative thermodynamic geometry of nonextensive ideal Classical, Bose and Fermi gases.We show that the intrinsic statistical interaction of nonextensive Bose (Fermi) gas is attractive (repulsive) similar to the extensive case but the value of thermodynamic curvature is changed by nonextensive parameter. In contrary to the extensive ideal classical gas, the nonextensive one may be divided to two different regimes. According to deviation parameter of the system to the nonextensive case, one can find a special value of fugacity, $z^{*}$, where the sign of thermodynamic curvature is changed. Therefore, we argue that the nonextensive parameter induces an attractive (repulsive) statistical interaction for $z<z^{*}$ ($z>z^{*}$) for an ideal classical gas. Also, according to the singular point of thermodynamic curvature, we consider the condensation of nonextensive Boson gas.

\end{abstract}
\maketitle

\section{Introduction}
A fundamental principle of quantum statistical mechanics is the existence of two types of particles, bosons
and fermions, obeying Bose-Einstein(BE) and Fermi-Dirac(FD) statistics. The Pauli exclusion principle dictates that no two fermions can occupy the same quantum state, while there is no such restriction on bosons\cite{Gasior}.

The fractional quantum Hall effect is one of the phenomena that forced physicists to regard some generalizations to the ordinary BE and FD statistics\cite{Laughlin}. The generalized Pauli principle or fractional exclusion statistics introduced by Haldane \cite{Haldane}, is one of the popular generalizations. A fractional parameter, $g$, is defined for fractional exclusion statistics. This parameter is related to the change of the dimension of the single particle Hilbert space with respect to the change of the number of particles, when the size of the system and the boundary conditions are unchanged. By definition, also, $g=0(g=1)$ corresponds to bosons(fermions).

Fractional exchange statistics is another concept of fractional statistics that is defined when the many body wave function of a system of indistinguishable particles is allowed to have an arbitrary phase $e^{i \pi \alpha}$ upon an adiabatic exchange process of two particles. Here, $\alpha$ is the so-called exchange statistical parameter, interpolating between $\alpha= 0$ (bosons) and $\alpha = 1$ (fermions). Fractional exchange statistics is limited to two spatial dimensions. However, fractional exclusion statistics, is based on the structure of the Hilbert space, rather than on the configuration space, of the particle assembly and is, thus, not restricted to $d \leqslant 2$ \cite{Pellegrino,Murthy}.

Another types of fractional exclusion statistics are Polychronakos \cite{Polychronakos} and Gentile statistics \cite{Gena,Genb}, so the latter is the generalization of Fermi-Dirac and Bose-Einstein statistics which is based on allowing particles to occupy the same quantum states.

Furthermore, there is another generalization of BE and FD statistics based on nonextensive entropy or q-generalized entropy ($S_q$), introduced by Tsallis \cite{Tsallisc} as a generalization of Boltzmann-Gibbs (BG) entropy.
The nonextensive entropy, $S_q$ and it's associated statistical mechanics have diverse applications including solar wind systems \cite{Burlagaa,Burlagab}, quantum entangled and nonentangled systems \cite{Caruso,Saguia,Nobre,Nayak}, cold atoms \cite{Douglas,Bagci,Lutz}, plasmas \cite{Liu,Ghosh,Guo}, trapped atoms \cite{Devoe}, spin-glasses\cite{Pickup}, black holes and cosmology \cite{Oliviera,Komatsu}, economics \cite{Borland,Ludescher}, earthquakes \cite{Antonopoulos}, self-organized criticality \cite{Tamarit}, etc.

In current paper we will investigate the q-generalized BE, FD and Maxwell-Boltzmann (MB) statistics through the thermodynamic geometry. The thermodynamic curvature has already been calculated for many various models\cite{Ruppeinerb,Brody,Ruppeinerc,Ruppeiner,Weinhold,Janyszek,Mirza16,Ruppeiner16,Ruppeiner15}. The thermodynamic curvature of the ideal classical gas is zero and it could be a criterion for statistical interaction of the system \cite{Ruppeiner,Weinhold}. Janyszek and Mrugala \cite{Janyszek} worked out the thermodynamic curvature of the ideal Fermi and Bose gases and reported that the sign of the thermodynamic curvature always is different for ideal Fermi and Bose gases. It may be shown that the sign of thermodynamic curvature specifies the attractive or repulsive statistical interaction of the systems. It has been argued that the scalar curvature could be used to show that fermion gases are more stable than boson gases. Also, phase transition properties of van der Waals gas and some other thermodynamic models have been considered and it has been shown that the singular point of the thermodynamic curvature coincides with the critical point of the system \cite{Brodyb,Janke}. Recently, in a series of articles, the thermodynamic curvature of some intermediate statistics such as Haldane fractional exclusion statistics, Polychronakos statistics, Gentile statistics, q-deformed boson and fermion have been investigated \cite{Mirza1,Mirza2,Mirzaa,Mirzab,Mirza4}. Also, some applications have been represented for these statistics \cite{Mohammadzadeh1,Mohammadzadeh2}.

The paper is organized as follows. In Sec.II, we briefly introduce the nonextensive entropy. In Sec.III, we present a short review on the thermodynamic geometry and its evaluation for the well-known ideal qunatum and classical gases. In Sec.IV, we evaluate the thermodynamic quantities of the nonextensive BE, FD and classical systems for small deviation from the extensive cases. Also, we work out the thermodynamic curvature of these systems. In Sec.V, we focus on the phase transition point of the nonextensive BE statistics and obtain the phase transition temperature. Finally, we conclude the paper in Sec.VI.
\section{Nonextensive Entropy}\label{1}
The goal of statistical mechanics is reaching the thermodynamic relations, starting from the microscopic rules (classical, relativistic, quantum mechanics, chromodynamics) and using probability theory. Along the connections between the macroworld and microworld, the ultimate link is made through the fundamental concept of entropy \cite{Gibbs}. It is clear that Boltzmann-Gibbs (BG) entropy satisfactorily describes the nature if the effective microscopic interactions and memory are short-ranged and the boundary conditions are non(multi)fractal\cite{Tsallisa}. The BG entropy of the commonly known thermodynamical systems is extensive. This important property means that the entropy of the system is proportional to the size of the system. However, the BG entropy of some systems with Long-range interparticle interaction, long-term microscopic or mesoscopic memory, fractal or multifractal occupation in phase space, cascade transfer of energy or information, and intrinsic fluctuations of some dynamical system parameters is nonextensive and some kind of generalization appears to become necessary \cite{Budini}.

For an important class of such systems, there exists an entropy which is extensive in terms of the microscopic probabilities \cite{Tsallisb}.
The additive BG entropy can be generalized into the nonadditive q-entropy \cite{Tsallisc}
\begin{equation}\label{eq1}
 S_q \equiv-k_B\frac{1-\sum_{i=1}^W p_i^q}{1-q};  q \in  \mathbf{R},
\end{equation}
where, $S_{q=1}=S_{BG}\equiv-k_B\sum_{i=1}^W p_i \ln {p_i}$. Also, $p_i$ is the probability of finding the system in the microscopic state $i$, $k$ is the Boltzmann constant, $W$ is the total number of microstates and real parameter $q$ determines the degree of nonadditivity.
This is the basis of the so called nonextensive statistical mechanics \cite{Tsallisbook}, which generalizes the BG entropy.

The generalized forms of the distribution functions have been found using Boltzmann’s H-theorem where the entropy must be maximum. The validity of H-theorem for the generalized entropies, specially for the Tsallis entropy has been verified\cite{Ramshaw}.  The q-distribution function is given by \cite{Demirhan}
\begin{equation}\label{eq2}
n_q(\epsilon)=\frac{1}{{(1+(q-1)\beta(\epsilon-\mu))}^{\frac{1}{q-1}}+\alpha},
\end{equation}
where, $\beta=1/k_BT$, $\epsilon$ is the state energy of a particle, $\mu$ is the chemical potential and $\alpha=-1(\alpha=1)$ corresponds to bosons and fermions, while $\alpha=0$ corresponds to the classical particles which follows the Maxwell-Boltzmann(MB) statistics. The Eq.(\ref{eq2}) reduced to the well- known BE, FD and MB distribution functions for $\alpha=-1,1$ and $0$ at the limit of $q\rightarrow1$as follows
\begin{equation}\label{eq3}
n(\epsilon)=\frac{1}{e^{\beta(\epsilon-\mu)}+\alpha}.
\end{equation}
\section{Phase transitions in nonextensive systems}\label{3}
Experiences tells us that  the free energy for a large system ($\Omega$) is extensive, i.e. $F_\Omega \varpropto V(\Omega)$, where $V$ denotes the volume of the system. Thus, we can write \cite{Goldenfeld}:
\begin{equation}\label{eq17}
f_b [K] \equiv  \lim_{V(\Omega) \rightarrow \infty}\frac{F_\Omega}{V(\Omega)},
\end{equation}
where, $f_b$ is the bulk free energy per unit volume. Note that $[K] \equiv [\{ K_i \}]$ is a set of the coupling constants in the Hamiltonian of the system. For a system defined on a lattice, with $N(\Omega)$ lattice site, $N(\Omega)$ is substituted by $V(\Omega)$ as follows:
\begin{equation}\label{eq18}
f_b [K] \equiv  \lim_{N(\Omega) \rightarrow \infty}\frac{F_\Omega}{N(\Omega)}.
\end{equation}

The bulk free energy $f_b$ describes extensive thermodynamic behavior, but does not describe surface or finite size behavior. This information may be computed from the surface free energy. The limit in Eqs.(\ref{eq17}) and (\ref{eq18}) is known as the thermodynamic limit that is not trivial. In order for a uniform bulk behavior to exist, the thermodynamic limit must be taken carefully.

However, when the phase boundaries exist, phase transitions must come into two classes:

    (1) First order phase transition, where $\partial f_b / \partial K_i$ is discontinuous across a phase boundary

    (2) Second order phase transition or continuous phase transition,  where $\partial f_b / \partial K_i$ is continuous across a phase

          boundary.

The phase transition for the nonextensive systems was investigated by Gross in \cite{Grossa,Grossb,Grossc}, and it has been shown that the nonextensive systems do not allow to go to the thermodynamic limit. There, it was shown that the nonextensivity of inhomogeneous systems with separated phases gives just a clue to illuminate the physics of phase transitions explicitly and sharply. Gross has showed that canonical and grandcanonical ensembles can be deduced from microcanonical ensemble only
in the thermodynamic limit if the system is homogeneous, because if the thermodynamic limit does not exist, fluctuations of energy per particle do not vanish.

We can define the entropy (e.g. $S=K_B \ln W$) as a measure of the mechanical N-body phase space. So, thermodynamics has thus a geometrical interpretation and can be read off from the topology of $W(E,N,...)$, the volume of its constant energy manifold. On the other hand, that the micro-canonical statistics works well also for “small” systems
without invoking extensivity will be demonstrated here for finite normal systems which are also nonextensive at phase transitions of first order. Also, he used the most natural extension of thermo-statistics to many nonextensive systems without invoking any modification of the entropy like that proposed by Tsallis. Finally, Gross has presented that there is a "zoo" of phase transitions containing first order, second order and a multi-critical point in the nonextensive systems. However, we use the modification of the entropy and resign the Gross formalism for future works.
\section{Thermodynamic Geometry}\label{2}
The geometrical structure of phase space of statistical thermodynamics was introduced by Gibbs \cite{Gibbs}. Then Ruppeiner and Weinhold developed the geometrical thermodynamics \cite{Ruppeiner,Weinhold}. They introduced two types of Riemannian metric structure representing thermodynamic fluctuation theory, which were connected to the second derivative of entropy or internal energy. The theory represents a new qualitative tool for the study of fluctuation phenomena.

Ruppeiner geometry is based on the entropy representation, where the extended set of $n + 1$ extensive variables of the system are denoted by $X=(U,N^1,\ldots,V,\ldots,N^\tau)$, while Weinhold geometry uses the energy representation in which the extended set of $n + 1$ extensive variables of the system are presented by
$Y = (S,N^1,\ldots,V,\ldots,N^\tau)$ . It may be noted that one can work in any thermodynamic potential representation that is the Legendre transform of the entropy or the internal energy. One can write the metric of this representation by the second derivative of the thermodynamic potential with respect to intensive variables; for example, the thermodynamic potential $\Phi$ which is defined as,
\begin{equation}\label{eq4}
\Phi=\Phi({\{F^i\}})=\Phi(1/T,−\mu^1/T, . . . , P/T, . . . ,−\mu^\tau/T),
\end{equation}
$\Phi$ is the Legendre transform of entropy with respect to the intensive parameter, $X^i$ and $\tau$ denotes the number of various types of particles. The metric in this representation is
\begin{equation}\label{eq5}
g_{ij}=\frac{\partial^2 \Phi}{\partial F^i \partial F^j}, ~~~~ F^i=\frac{\partial S}{\partial X^i} .
\end{equation}
Janyszek and Mrugala used the partition function to express the metric geometry of the parameter space \cite{Janyszek},
\begin{equation}\label{eq6}
g_{ij}=\frac{\partial^2 \ln {Z}}{\partial \beta^i \partial \beta^j},
\end{equation}
where, $\beta^i = F^i/ k$ and $Z$ is the partition function. The two dimensional parameter space is defined by
\begin{equation}
\beta^1=\beta=\frac{1}{kT},~~\beta^2=\gamma=-\frac{\mu}{kT}.
\end{equation}
So we can write
\begin{align}\label{eq7}
&g_{\beta \beta}=\frac{\partial^2 \ln {Z}}{\partial \beta^2}=-{(\frac{\partial U}{\partial \beta})}_\gamma,\nonumber\\
&g_{\beta \gamma}=g_{\gamma \beta }=\frac{\partial^2 \ln {Z}}{\partial \beta \partial \gamma}=-{(\frac{\partial U}{\partial \gamma})}_\beta,\nonumber\\
&g_{\gamma \gamma}=\frac{\partial^2 \ln {Z}}{\partial \gamma^2}=-{(\frac{\partial N}{\partial \gamma})}_\beta.
\end{align}
In two dimensional spaces, the formula for $R$ may be written as \cite{Janyszek}
\begin{equation}\label{eq8}
R=-\frac{2 \begin{vmatrix}
g_{\beta \beta}& g_{\beta \gamma} & g_{\gamma \gamma}\\
 g_{\beta \beta,\beta}& g_{\beta \gamma,\beta} & g_{\gamma \gamma,\beta}\\
g_{\beta \beta,\gamma}& g_{\beta \gamma,\gamma} & g_{\gamma \gamma,\gamma}

\end{vmatrix}}{{\begin{vmatrix}
g_{\beta \beta}& g_{\beta \gamma}\\
g_{\beta \gamma}&g_{\gamma \gamma}
\end{vmatrix}}^2},
\end{equation}
where
\begin{align}\label{eq9}
& g_{\beta \beta,\beta}=\frac{\partial g_{\beta \beta}}{\partial \beta}, \nonumber \\
& g_{\beta \beta,\gamma}=g_{\beta \gamma, \beta}=\frac{\partial g_{\beta \gamma}}{\partial \beta},\nonumber\\
& g_{\gamma \gamma,\beta}=g_{\beta \gamma, \gamma}=\frac{\partial g_{\gamma \gamma}}{\partial \beta},\nonumber\\
& g_{\gamma \gamma,\gamma}=\frac{\partial g_{\gamma \gamma}}{\partial \gamma}.
\end{align}
Now, we can construct the thermodynamic geometry of any thermodynamical systems. In the following, we focus on the quantum and classical ideal gas. In the thermodynamic limit, the internal energy and particle number of an ideal gas in a $D$ dimensional box of volume $L^D$ with the dispersion relation
\begin{equation}\label{eq10}
\epsilon =a{p}^{\sigma},
\end{equation}
can be written as
\begin{align}\label{eq11}
&U=\int_0^{\infty}\epsilon n(\epsilon) \Omega(\epsilon) \mathrm{d} \epsilon, \nonumber \\
&N=\int_0^{\infty} n(\epsilon) \Omega(\epsilon) \mathrm{d} \epsilon,
\end{align}
and $\Omega(\epsilon)$ is the density of the single particle state for the system. Neglecting the spin
of particle, the standard form of density of states will be
\begin{equation}\label{eq12}
\Omega(\epsilon)=\frac{A^D}{\Gamma({\frac{D}{2}})}\epsilon^{D/ \sigma-1},
\end{equation}
where, $\sigma$ takes some values for example ($\sigma= 2$) for non relativistic and ($\sigma = 1$) for ultra
relativistic particles and $A=L\sqrt{\pi}/(a^{1/\sigma}h)$ is a constant and for simplicity we will set it equal to one ($A=1$).

According to the mean occupation number of quantum Bose and Fermi particles and the classical particles, we can evaluate internal energy and particle number using Eq. (\ref{eq11}).

\begin{figure}[h]
\centerline{\includegraphics[scale=0.45]{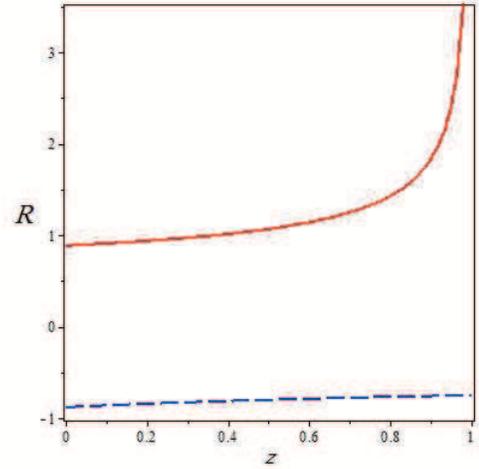}}
\caption{Thermodynamic curvature of an ideal Bose gas ($R_{BE}$) [$solid line$]  and an ideal Fermi gas ($R_{FD}$) [$dashed line$] versus fugacity($z$).}
\label{fig1}
\end{figure}

For an ideal Bose and Fermi gas, we obtain the internal energy and particle number as follows
\begin{align}\label{eq13}
&U_{\substack {BE \\ FD}}=\pm \frac{\beta^{-5/2}}{\Gamma(\frac{3}{2})}\Gamma({\frac{5}{2}})g_{5/2}(\pm z), \nonumber \\
&N_{\substack {BE \\ FD}}=\pm \beta^{-3/2} g_{3/2}(\pm z),
\end{align}
where
\begin{equation}\label{eq14}
g_\nu (z)=\frac{1}{\Gamma (\nu)} \int ^ \infty_0 \frac{x^{\nu-1}dx}{z^{-1}e^x-1}=z+\frac{z^2}{2^\nu}+\frac{z^3}{3^\nu}+\ldots
\end{equation}
denotes the Bose-Einstein functions, so that $z=\exp (\mu/k_B T)$ is the fugacity of the gas, where, $\mu$ is the chemical potential. After some calculations, the thermodynamic curvature for the distribution function presented in Eq. (\ref{eq3}) becomes
\begin{equation}\label{eq15}
R_{\substack {BE \\ FD}}=-\frac{2 \begin{vmatrix}
\pm \frac{15}{4}\frac{g_{5/2}(\pm z)}{\beta^{7/2}}&\pm\frac{3}{2}\frac{g_{3/2}(\pm z)}{\beta^{5/2}} & \pm\frac{g_{1/2}(\pm z)}{\beta^{3/2}}\\
\mp\frac{105}{8}\frac{g_{5/2}(\pm z)}{\beta^{9/2}}&\mp \frac{15}{4}\frac{g_{3/2}(\pm z)}{\beta^{7/2}} & \mp\frac{3}{2}\frac{g_{1/2}(\pm z)}{\beta^{5/2}}\\
\mp\frac{15}{4}\frac{g_{3/2}(\pm z)}{\beta^{7/2}} &  \mp\frac{3}{2}\frac{g_{1/2}(\pm z)}{\beta^{5/2}} & \mp\frac{g_{-1/2}(\pm z)}{\beta^{3/2}}

\end{vmatrix}}{{\begin{vmatrix}
\pm \frac{15}{4}\frac{g_{5/2}(\pm z)}{\beta^{7/2}}&\pm \frac{3}{2}\frac{g_{3/2}(\pm z)}{\beta^{5/2}}\\
\pm \frac{3}{2}\frac{g_{3/2}(\pm z)}{\beta^{5/2}}&\pm \frac{g_{1/2}(\pm z)}{\beta^{3/2}}
\end{vmatrix}}^2}.
\end{equation}
In Fig.(\ref{fig1}) we represent $R_{BE}$ and $R_{FD}$ versus $z$ for $\beta=1$ since it is counted in just a coefficient. It is clear that we must take $z<1$ in the Bose gas, as regards Bose-Einstein condensation takes place at $z=1$. We denote that $R$ is positive for Bose gas and is negative for Fermi gas, so that shows interactions between bosons is attractive and between fermions is repulsive.

For an ideal classical gas, we obtain the internal energy and particle number as below
\begin{align}\label{eq16}
&U_{MB}=\frac{3}{2}z \beta^{-5/2}, \nonumber \\
&N_{MB}=z \beta^{-3/2}.
\end{align}
The thermodynamic curvature becomes zero for this case($R_{MB}=0$).

\section{Thermodynamic curvature of $q$-generalized distribution functions}\label{4}
Analytical investigation of q-generalized statistics for all values of $q$ is not possible because of the complicated form of the q-generalized distribution function. In this paper, we consider small deviation of nonextensive distribution function with respect to the extensive one.

The first order approximation of q-generalized distribution function can be evaluated as follows \cite{Tirnakli}:
\begin{equation}\label{eq19}
n(\epsilon)=\frac{1}{e^{\beta(\epsilon-\mu)}+\alpha}+\frac{1}{2}e^{\beta(\epsilon-\mu)}\bigg(\frac{\beta{(\epsilon-\mu)}}{(e^{\beta(\epsilon-\mu)}+\alpha)}\bigg)^2 r+O(r^2)
\end{equation}
where, $r=q-1$ is a small deviation parameter with respect to the standard distributions. In the following, we work out the thermodynamic curvature of the q-generalized BE, FD and MB statistics.
\subsection{q-generalized BE (FD) statistics}
\begin{figure}
\centerline{\includegraphics[scale=0.45]{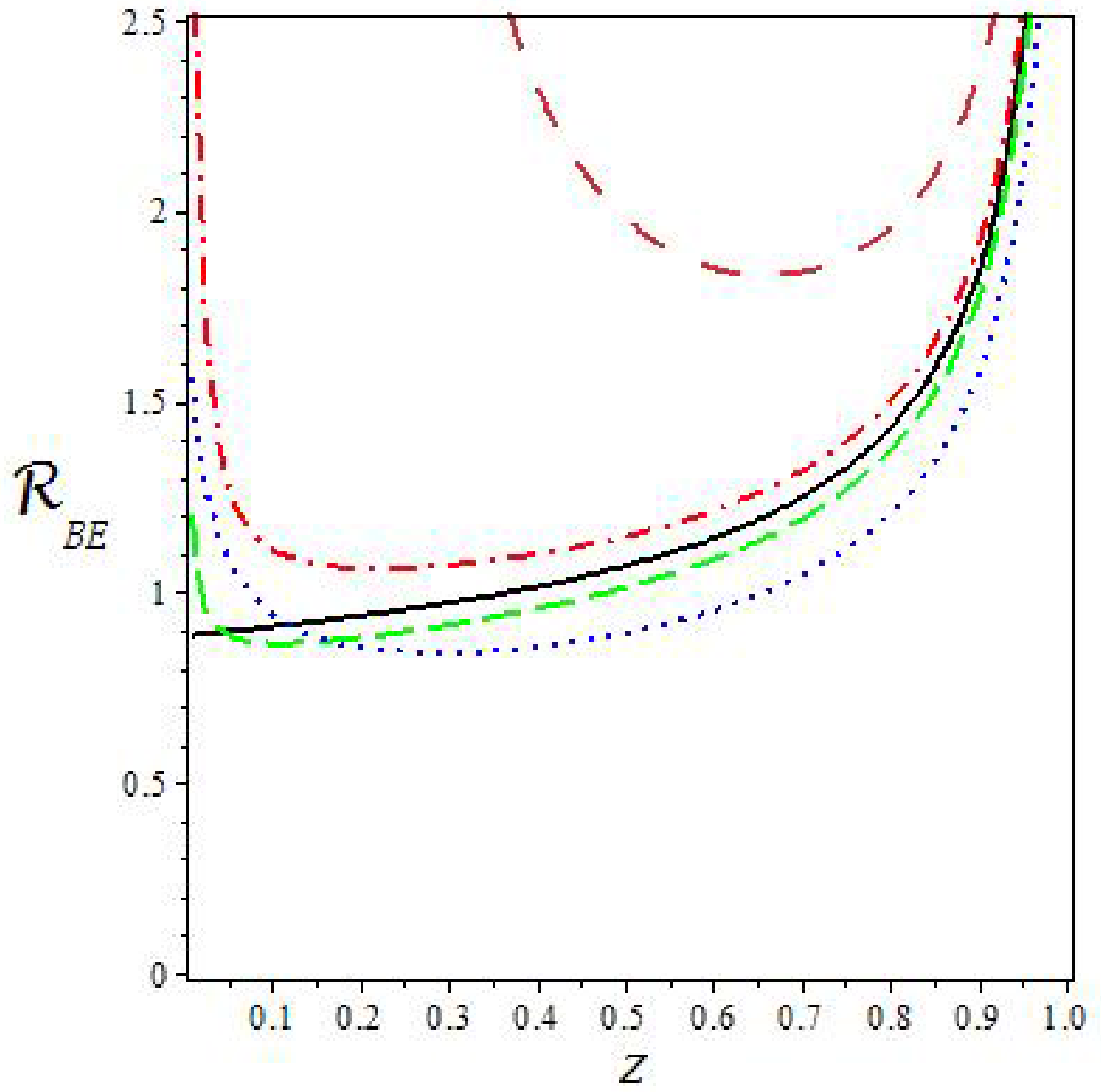}}
\caption{Thermodynamic curvature of an approximated q-generalized BE system versus fugacity $z$ for some values of $r=0$[$solid line$],0.01[$dashed line$], -0.01 [$dash-dotted line$] , 0.05 [$dotted line$] and -0.05[$space-dashed line$].}
\label{fig2}
\end{figure}

\begin{figure}
\centerline{\includegraphics[scale=0.45]{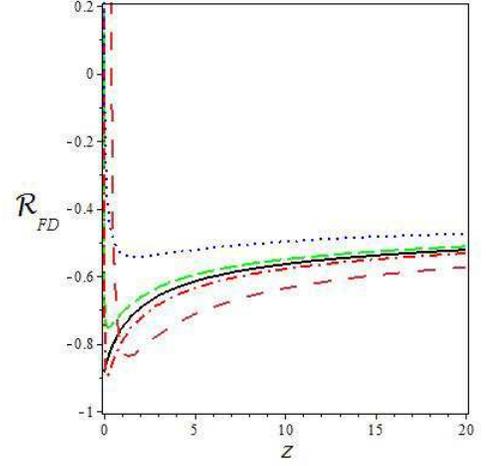}}
\caption{Thermodynamic curvature of an approximated q-generalized FD system versus fugacity $z$ for some values of $r=0$[$solid line$],0.01[$dashed line$], -0.01 [$dash-dotted line$] , 0.05 [$dotted line$] and -0.05[$space-dashed line$].}
\label{fig3}
\end{figure}

With regards to Eq.(\ref{eq19}), we obtain the first order approximation distribution function of q-generalized BE(FD) statistics as follows:
\begin{equation}\label{eq20}
n(\epsilon)\simeq \frac{1}{e^{\beta(\epsilon-\mu)}\pm 1}+\frac{1}{2}e^{\beta(\epsilon-\mu)}\bigg(\frac{\beta{(\epsilon-\mu)}}{e^{\beta(\epsilon-\mu)}\pm1}\bigg)^2 r.
\end{equation}
Now, using Eq. (\ref{eq11}) we evaluate the thermodynamic quantities such as
\begin{align}\label{eq21}
&U= U_{\substack {BE \\ FD}}+r \mathcal{U}_{\substack {BE \\ FD}}, \nonumber \\
&N=N_{\substack {BE \\ FD}}+r \mathcal{N}_{\substack {BE \\ FD}},
\end{align}
where
\begin{align}\label{eq22}
\mathcal{U}_{\substack {BE \\ FD}}=\pm\frac{\beta^{-5/2}}{\Gamma(\frac{3}{2})}\bigg(&\frac{7}{4}\Gamma({\frac{7}{2}})g_{7/2}(\pm z)- \frac{5}{2} \ln(z)\Gamma({\frac{5}{2}})g_{5/2}(\pm z)\nonumber\\&+\frac{3}{4}{\ln(z)}^2\Gamma({\frac{3}{2}})g_{3/2}(\pm z)\bigg), \nonumber\\
\mathcal{N}_{\substack {BE \\ FD}}=\pm \frac{\beta^{-3/2}}{\Gamma(\frac{3}{2})}\bigg(&\frac{5}{4}\Gamma({\frac{5}{2}})g_{5/2}(\pm z)-\frac{3}{2} \ln(z)\Gamma({\frac{3}{2}})g_{3/2}(\pm z)\nonumber\\&+\frac{1}{4}{\ln(z)}^2\Gamma({\frac{1}{2}})g_{1/2}(\pm z)\bigg).
\end{align}
It is clear that the metric elements of approximated q-generalized BE(FD) have the form
\begin{equation}\label{eq23}
G_{\mu \nu}=g_{\mu \nu}+r \eta_{\mu \nu},
\end{equation}
where, $g_{\mu \nu}$ has been defined previously in Eq.(\ref{eq7}) and $\eta_{\mu \nu}$ is obtained as follows
\begin{align}\label{eq24}
\eta_{\beta \beta}=&-{(\frac{\partial \mathcal{U}_{\substack {BE \\ FD}}}{\partial \beta})}_\gamma=\pm \frac{15}{32\beta^{7/2}}\bigg[ 35g_{7/2}(\pm z)\nonumber\\&-20 \ln(z)g_{5/2}(\pm z)+4 {\ln(z)}^2   g_{3/2}(\pm z) \bigg], \nonumber \\
\eta_{\beta \gamma}=&-{(\frac{\partial \mathcal{U}_{\substack {BE \\ FD}}}{\partial \gamma})}_\beta=\pm \frac{3}{16\beta^{5/2}}\bigg[ 15g_{5/2}(\pm z)\nonumber\\&-12 \ln(z)g_{3/2}(\pm z)+4 {\ln(z)}^2   g_{1/2}(\pm z) \bigg], \nonumber \\
\eta_{\gamma \gamma}=&-{(\frac{\partial \mathcal{N}_{\substack {BE \\ FD}}}{\partial \gamma})}_\beta=\pm \frac{1}{8\beta^{3/2}}\bigg[ 3g_{3/2}(\pm z)\nonumber\\&-4 \ln(z)g_{1/2}(\pm z)+4 {\ln(z)}^2   g_{-1/2}(\pm z) \bigg].
\end{align}
Consequently, we can write the new form of $R$ as
\begin{equation}\label{eq25}
\mathcal{R}_{\substack {BE \\ FD}}=-\frac{2 \begin{vmatrix}
G_{\beta \beta}& G_{\beta \gamma} & G_{\gamma \gamma}\\
 G_{\beta \beta,\beta}& G_{\beta \gamma,\beta} & G_{\gamma \gamma,\beta}\\
G_{\beta \beta,\gamma}& G_{\beta \gamma,\gamma} & G_{\gamma \gamma,\gamma}

\end{vmatrix}}{{\begin{vmatrix}
G_{\beta \beta}& G_{\beta \gamma}\\
G_{\beta \gamma}&G_{\gamma \gamma}
\end{vmatrix}}^2},
\end{equation}
or
\begin{equation}\label{eq26}
\mathcal{R}_{\substack {BE \\ FD}}=R_{\substack {BE \\ FD}}+f(r,g_{\mu \nu},\eta_{\mu \nu},g_{\mu \nu, \gamma},\eta_{\mu \nu,\gamma}).
\end{equation}
We plot the thermodynamic curvature of an approximated q-generalized BE (FD) statistics versus fugacity, $z$,  for different values of $r$ Fig.(\ref{fig2}) (Fig.(\ref{fig3})).
\subsection{q-generalized MB statistics}
\begin{figure}
\centerline{\includegraphics[scale=0.45]{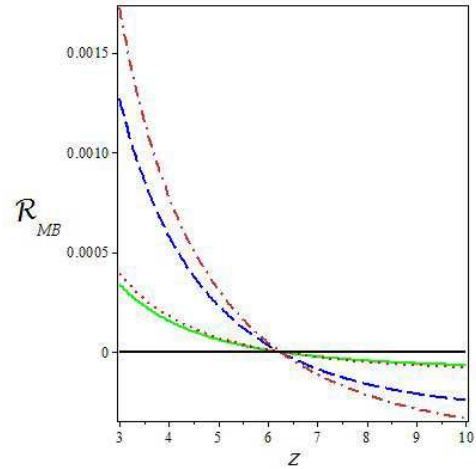}}
\caption{Thermodynamic curvature of an approximated q-generalized MB system versus fugacity $z$ for some values of $r=0.01[solid line], -0.01 [dotted line] , 0.02 [dashed line] and -0.02[dash-dotted line]$ .}
\label{fig4}
\end{figure}

Along the previous subsection, we do the same calculation for q-generalized MB statistics. The internal energy and particle number of MB statistics for small deviation parameter of standard MB distribution can be written as
\begin{align}\label{eq27}
&U= U_{MB}+r \mathcal{U}_{MB}, \nonumber \\
&N=N_{MB}+r \mathcal{N}_{MB},
\end{align}
where
\begin{align}\label{eq28}
\mathcal{U}_{MB}=z \bigg( \frac {105}{16}-\frac {15}{4}\ln(z) +\frac{3}{4} \ln(z)^2 \bigg) \beta^{-5/2}, \nonumber\\
\mathcal{N}_{MB}=z \bigg( \frac {15}{8}-\frac{3}{2}\ln(z)+\frac{1}{2}\ln(z)^2 \bigg) \beta^{-3/2}.
\end{align}
Then we write the metric elements with differentiation of U and N in Eq.(\ref{eq27}) as follows:
\begin{align}\label{eq29}
G_{\beta \beta}&=-{(\frac{\partial U}{\partial \beta})}_\gamma \nonumber\\&=\frac{15z}{32\beta^{7/2}}\bigg[ 8+r\bigg(35-20\ln(z)+4\ln(z)^2 \bigg) \bigg], \nonumber \\
G_{\beta \gamma}&=-{(\frac{\partial U}{\partial \gamma})}_\beta\nonumber\\&=\frac{3z}{16\beta^{5/2}}\bigg[ 8+r\bigg(15-12\ln(z)+4\ln(z)^2 \bigg) \bigg], \nonumber \\
G_{\gamma \gamma}&=-{(\frac{\partial N}{\partial \gamma})}_\beta \nonumber\\&=\frac{1z}{8\beta^{3/2}}\bigg[ 8+r\bigg(3-4\ln(z)+4\ln(z)^2 \bigg) \bigg].
\end{align}
Thermodynamic curvature of nonextensive MB statistics has the explicit form as follows:
\begin{equation}\label{eq30}
\mathcal{R}_{MB}=\frac{A}{B},
\end{equation}
where
\begin{align}\label{eq31}
A=1280\,\beta^{3/2}r^2 \bigg( &-48 \ln(z) +90 r\,\ln(z) -60 r \ln(z) ^2\nonumber\\&+8 r \ln(z)^3-15 r+88 \bigg),\nonumber\\
B=z \bigg( 64+400 r&-192 r \ln(z) +64 r \ln(z)^2-75r^2 \nonumber\\&+40 r^2 \ln(z) +184 r^2 \ln(z)^2\nonumber\\&-96 r^2 \ln(z)^3+16 r^2 \ln(z) ^4 \bigg) ^2.
\end{align}
If we expand the Eq.(\ref{eq30}) with respect to $r$, it has the following form:
\begin{equation}\label{32}
\mathcal{R}_{MB}=\frac{5 \beta^{3/2}}{16 z} \bigg(-48 \ln(z)+88 \bigg) r^2+O(r^3).
\end{equation}
From the above equation we find that there is a special value of $z=z^*$ wherein the thermodynamic curvature is zero. This result shows that q-generalized MB statistics remains at classical state within a point related to a special fugacity . We plot  the thermodynamic curvature of an approximated q-generalized MB statistical system versus fugacity for different values of $r$ in Fig.(\ref{fig4}). It is obvious that the value of thermodynamic curvature is little, but for $z<z^*$, it is positive while for $z>z^*$, it is negative. It means that the q-generalized MB ideal gas has attractive (repulsive) intrinsic statistical interaction for $z<z^*$ ($z>z^*$).
\section{Condensation of Nonextensive Bose gas}\label{5}
For an ideal Bose gas ($q=1$), a phase transition point at $z=1$ is well-known and so-called Bose-Einstein condensation. Also, the thermodynamic curvature of this system is singular at $z=1$ according to Fig. (\ref{fig1}). A similar behavior is observed from  Fig. (\ref{fig2}) for all values of $r$ for an ideal q-generalized Bose gas. In the following, we explore the phase transition temperature which could be dependent on the value of $q$.

We evaluate the critical temperature using Eq. (\ref{eq21}), where we set $z=1$ at condensation point:
\begin{equation}\label{eq33}
T_c^q=\frac{h^2}{2 \pi mk_B}{\bigg\{\frac{N}{V\Big( \zeta(\frac{3}{2})+\frac{15}{8}r \zeta(\frac{5}{2})\Big)}\bigg\}}^ \frac{2}{3},
\end{equation}
and
 \begin{equation}\label{eq34}
\frac{ T_c^q}{T_c^{BE}}={\bigg\{  \frac{\zeta(\frac{3}{2})}{\zeta(\frac{3}{2})+\frac{15}{8}r \zeta(\frac{5}{2})}   \bigg\}}^{2/3}\simeq 1-\frac{5}{4} \frac{\zeta(\frac{5}{2})}{\zeta(\frac{3}{2})} r,
\end{equation}
where, $T_c^{BE}$ is the condensation temperature of an ideal Bose gas($r=0$). For any deviation from the standard BE distribution, the phase transition temperature can be increased (decreased) for $q<1 (q>1)$ with respect to Bose-Einstein condensation temperature.

\begin{figure}
\centerline{\includegraphics[scale=0.45]{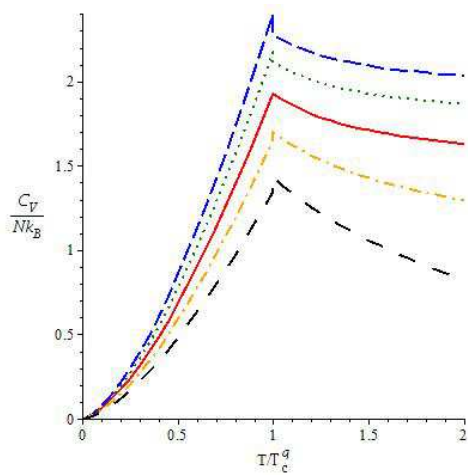}}
\caption{The specific heat of a nonextensive Bose gas as a function of the normalized temperature $T/T_c^q$ for some values of $r=0$[$solid line$], 0.01 [$dotted line$] , 0.02 [$dashed line$], -0.01[$dash$-$dotted line$] and -0.02[$space-dashed line$]. }
\label{final}
\end{figure}

Furthermore, regarding Eqs. (\ref{eq21}) and (\ref{eq22}) and using the well-known relation between the pressure, $P$, and the internal energy, $U$, \cite{Pathria}
\begin{equation}\label{eq35}
U = k_B T^2  \bigg\{ \frac{\partial}{\partial T} \bigg( \frac{PV}{k_B T} \bigg) \bigg\}  _{z,V}.
\end{equation}
We find the pressure as follows:
\begin{align}\label{eq36}
\frac{P}{k_B T}=\frac{1}{\lambda ^3} \bigg\{ &g_{5/2}(z)+r \bigg[ \frac{35}{8} g_{7/2}(z) \nonumber\\& -\frac{5}{2}\ln(z)g_{5/2}(z)+\frac{1}{2} \ln(z)^2 g_{3/2}(z)   \bigg] \bigg\},\nonumber\\
\end{align}
where $\lambda=h/(2 \pi mk_B T)^{\frac{1}{2}}$ is the thermal wavelength.
Now, for $T < T_c^q$, the pressure is given by Eq.(\ref{eq36}), with z replaced by unity:
\begin{equation}\label{eq37}
P(T)=\frac{k_B T}{\lambda ^3} \bigg[ \zeta (\frac{5}{2})+\frac{35}{8}r \zeta (\frac{7}{2}) \bigg],
\end{equation}
which is proportional to $T^{5/2}$ and is independent of $V$ and $N$. At the transition point the value of the pressure is
\begin{equation}\label{eq38}
P(T_c^q)=\bigg( \frac{2 \pi m}{h^2} \bigg)^{3/2} (k_BT_c^q)^{5/2} \bigg[ \zeta (\frac{5}{2})+\frac{35}{8}r \zeta (\frac{7}{2}) \bigg].
\end{equation}
Also, for $T > T_c^q$, the pressure is given by Eq.(\ref{eq36}), where $z(T)$ is determined by the implicit relationship
\begin{align}\label{eq39}
\frac{N}{V} \frac{h^3}{(2 \pi mk_B T)^{3/2}}=&g_{3/2}(z)+r \bigg[ \frac{15}{8} g _{5/2}(z)\nonumber\\&-\frac{3}{2} \ln(z) g_{3/2}(z)+\frac{1}{2} \ln(z)^2 g_{1/2}(z)\bigg]. \nonumber\\
\end{align}
Now, we are able to find the specific heat of the nonextensive Bose system from the thermodynamic relation
\begin{equation}\label{eq41}
\frac{C_V}{Nk_B}=\frac{1}{Nk_B}\bigg( \frac{\partial U}{\partial T} \bigg)_{N,V}=\frac{3}{2}\bigg[ \frac{\partial}{\partial T} \bigg( \frac{PV}{Nk_B} \bigg) \bigg]_\nu.
\end{equation}
We don't write the relations for $C_V$ explicitly due to it's complexity, but confine ourselves to the plot of the specific heat versus the temperature(Fig.(\ref{final})). It is obvious that the derivative of specific heat is discontinuous at $T=T_c^q$.
\section{Conclusion}\label{6}
The thermodynamic geometry of q-generalized BE, FD and MB statistics was explored. Also, Bose-Einstein condensation in the q-generalized BE system was investigated and the critical temperature of the q-generalized BE statistics was obtained. Finally, the specific heat of the BE statistical system was worked out. We note that all considerations is related to the small deviation from the standard distribution. Therefore, we have restricted ourselves to the first approximation limit.

From Fig. (\ref{fig2}), it was found that for all small values of nonextensivity parameter, $r=q-1$, the thermodynamic curvature, $R$, has a singularity at $z=1$, which implies that there is a condensation for all values of $r$. Also, we saw that the thermodynamic curvature is positive. Thus, the general intrinsic statistical interaction is attractive. For $r>0$ ($r<0$) the value of curvature is less (greater) than the curvature of standard ideal BE gas. We may argue that the deviation of the nonextensive parameter from the standard value makes the system more (less) stable for $r>0$ ($r<0$) with respect to the standard one ($r=0$). Some anomalies can be found in limit of $z=0$. It seems that the first order approximation breaks down at this limit and we need to consider it in more details by numerical methods. The non-perturbative investigation of q-generalized systems by numerical methods is in progress.

From the existence of the singularity of thermodynamic curvature at $z=1$, we argued that there is a phase transition. So, we obtained the critical temperature for the q-generalized BE statistics and compared with the ideal BE statistics one. It may be assume that the experimental value of the critical temperature is corresponding to the q-generalized BE statistics with a special value of $r$.

Also, we have plotted the diagram of the specific heat for some values of $r$. There is a discontinuity of the derivative of specific heat at $T=T_c^q$.

Also, we calculated the thermodynamic curvature of the q-generalized FD statistical systems and found that its sign is not changed for various $r$. There, the general intrinsic statistical interaction of q-generalized FD gas is repulsive. It is mentionable that the value of the curvature is less (greater) than the curvature of standard fermion gas for $r<0$ ($r>0$)

The same calculations were performed for the MB statistics. We saw that the thermodynamic curvature diagram versus fugacity represented an exciting result that is the existence of various signs of $R$ when $z$ is changed. For any value of $r$ there is a special value of fugacity, namely $z^*$, within that $R=0$ . At this point, the manner of the system is classical, but the sign of thermodynamic curvature is positive (negative) for $z<z^*$ ($z>z^*$). It means that the statistical interaction is attractive (repulsive) for  $z<z^*$ ($z>z^*$).


\end{document}